\documentclass[12pt,letterpaper]{article}
\usepackage{amsmath,amssymb}

\author{W.~M\"uck\thanks{E-mail: \texttt{wmueck@sfu.ca}}~\
and K.~S.~Viswanathan\thanks{E-mail: \texttt{kviswana@sfu.ca}}\\
\emph{\small Department of Physics, Simon Fraser University, Burnaby, B.C.,
V5A 1S6 Canada}}
\title{Counterterms for the Dirichlet Prescription of the AdS/CFT
Correspondence}

\begin{document}

\providecommand{\bvec}[1]{\ensuremath{\mathbf{#1}}}
\providecommand{\hR}{{\hat R}}
\providecommand{\e}{\mathrm{e}}
\providecommand{\kint}{\int\frac{d^dk}{(2\pi)^d}\,}
\providecommand{\eikx}{\e^{-i\bvec{k\cdot x}}}

\maketitle
\begin{abstract}
We illustrate the Dirichlet prescription of the AdS/CFT correspondence
using the example of a massive scalar field and argue that it is the only
entirely consistent regularization procedure known so far. Using the Dirichlet
prescription, we then calculate the
divergent terms for gravity in the cases $d=2,4,6$, which give rise to
the Weyl anomaly in the boundary conformal field theory.
\end{abstract}

\newpage
\section{Introduction}
\label{intro}
It has been stated in most papers on this subject that the
correspondence between a field theory on anti-de Sitter space (AdS)
and a conformal field theory (CFT) on its horizon is formally
described by the formula \cite{Gubser,Witten} 
\begin{equation}
\label{form}
  \int_{\phi_0} \mathcal{D}\phi\, \e^{-I_{AdS}[\phi]} 
  = \left\langle \exp \int d^dx\; \phi_0(x)
  \mathcal{O}(x) \right\rangle, 
\end{equation}
where the functional integral on the left hand side is over all fields
$\phi$, which asymptotically approach $\phi_0$ on the AdS horizon. 
On the right hand side, $\phi_0$ couples as a current to some boundary
conformal field $\mathcal{O}$. In the classical approximation the left
hand side is identical to $\exp(-I[\phi_0])$, where $I[\phi_0]$ is the
on-shell action evaluated as a functional of the boundary value. 
Thus, the formula \eqref{form} enables one to calculate correlation
functions of the field $\mathcal{O}$ in the boundary conformal field theory. 
This rather formal identification of partition functions needs
refinement due to the fact that $I[\phi_0]$ is divergent as a result
of the divergence of the AdS metric on the horizon.
Let us choose the conventional representation of anti de-Sitter space,
namely the upper half space $x^0>0$, $\bvec{x}\in\mathbb{R}^d$ with
the metric
\begin{equation}
\label{metric}
  ds^2 = \frac1{(x^0)^2} \left[(dx^0)^2 +(d\bvec{x})^2 \right].
\end{equation}
The horizon is given by $x^0=0$, but in order to regularize the action
one considers the space restricted to $x^0>\epsilon$. The regularized
on-shell action will be a function of $\epsilon$. Moreover, the terms which
diverge in the limit $\epsilon\to0$ can be isolated and cancelled with
counterterms. The remaining finite result is identified with the
right hand side of eqn.\ \eqref{form}. There is a subtlety
concerning the proper choice of boundary values, but consistency
forces us to use the boundary values at $x_0=\epsilon$ (We call this
the Dirichlet prescription). This subtlety
and its resolution is illustrated for the example of the massive
scalar field in section~\ref{prod}. 

The Dirichlet prescription of the AdS/CFT correspondence has been used
to successfully calculate the two-point functions of scalar fields
\cite{Gubser,Mueck1,Freedman1}, spinors \cite{Mueck2}, vector fields
\cite{Mueck2}, Rarita Schwinger fields \cite{Corely} and gravitons
\cite{Mueck3}. It must be noted that the subtlety mentioned above 
affects neither the finite terms in the two-point functions for massless
scalar and vector fields, gravitons, spinors and Rarita Schwinger
fields \cite{Witten,Henningson1,Liu,AVolovich,Koshelev}, nor higher
point correlators (cf.\ \cite{Freedman2} and references therein).

More recently, attention has been brought to the divergent
contributions, which have to be cancelled by counterterms
\cite{Henningson2,Henningson3,Nojiri1,Hyun,Balasu,Nojiri2,Gonzalez,Emparan,Nishi}. 
Of particular importance are terms, which are logarithmically divergent,
since those counterterms are not invariant under conformal or Weyl
scaling transformations. Hence, the presence of a logarithmic
divergence leads to a conformal or Weyl anomaly in the finite part of
the action. The Weyl anomaly has recently been calculated for the
cases $d=2,4,6$ \cite{Henningson2,Henningson3}. However, the authors
of these papers used a regularization, which does not consistently
address the subtlety mentioned above. Therefore, we present in
section~\ref{anomaly} the calculation of the divergent terms for free
gravity using the Dirichlet prescription. Our results for the
terms relevant to the Weyl anomaly in $d=2,4,6$ agree with those of
\cite{Henningson2,Henningson3}, but we regard this as a coincidence
particular to gravitons. Finally, we urge the
reader to consult the appendix for our notation and for a review of
the time slicing formalism, which is used in section~\ref{anomaly}.

\section{The Regularization Procedure}
\label{prod}
We illustrate the regularization procedure with the example of the
free massive scalar field, whose action is given by 
\begin{equation}
\label{action}
  I = \frac12 \int d^{d+1}x \sqrt{g} \left( D_\mu \phi D^\mu \phi +
  m^2 \phi^2 \right),
\end{equation}
and whose equation of motion with the metric \eqref{metric} is 
\begin{equation}
\label{eqnmot1}
  \left[ x_0^2 \partial_\mu \partial_\mu - x_0(d-1)\partial_0 -m^2
  \right] \phi=0.
\end{equation}
The solution of eqn.\ \eqref{eqnmot1}, 
which does not diverge for $x_0\to\infty$ is given in terms of the mode 
\[ x_0^\frac{d}2 \eikx K_\alpha(kx_0), \qquad
\text{where} \quad \alpha= \sqrt{\frac{d^2}4+m^2} \]
and $K_\alpha$ is a modified Bessel function. 
Let us isolate the leading behaviour for small $x_0$ by defining
\begin{equation}
\label{phihatdef}
  \phi(x) = x_0^{\frac{d}2-\alpha} \hat\phi(x).
\end{equation}
Then, $\hat\phi$ has a finite limit as $x_0$ goes to zero. 
However, one must take care to express the regularized on-shell 
action in terms of
the boundary value at $x_0=\epsilon$. This is easiest done by using 
\begin{equation}
\label{phihat}
  \hat\phi(x) = \left(\frac{x_0}\epsilon\right)^\alpha \kint \eikx
  \frac{K_\alpha(kx_0)}{K_\alpha(k\epsilon)} \phi_\epsilon(\bvec{k}),
\end{equation}
which satisfies $\hat\phi(\bvec{x},\epsilon)=\phi_\epsilon(\bvec{x})$.
Consider the regularized on-shell action, which is \cite{Mueck1}
\begin{equation}
\label{onshell}
  I(\epsilon) = -\frac12 \int d^dx\, \epsilon^{-2\alpha} \left[ \left(
  \frac{d}2 -\alpha \right) \phi_\epsilon^2
  + \epsilon \phi_\epsilon \left. \partial_0
  \hat\phi\right|_\epsilon \right]
\end{equation}
The first term on the right hand side is divergent and must be
cancelled with a counterterm. The second term might contain other
divergent terms, but also gives rise to the finite term \cite{Mueck1}
\begin{equation}
\label{Ifin}
  I_{fin} = -\alpha c_\alpha \int d^dx d^dy\, \frac{\phi_\epsilon(\bvec{x})
  \phi_\epsilon(\bvec{y})}{|\bvec{x-y}|^{d+2\alpha}},
\end{equation}
where $c_\alpha = \Gamma(d/2+\alpha)/[\pi^\frac{d}2 \Gamma(\alpha)].$  

On the other hand, there appears to be a slightly different, and in our view
not entirely consistent, prescription. Essentially, it expresses 
$\hat\phi$ in terms of the boundary value $\phi_0$ at $x_0=0$, which 
can be done by writing
\begin{equation}
\label{phihat2}
  \hat\phi(x) = \frac{2^{1-\alpha}}{\Gamma(\alpha)} \kint \eikx
  (kx_0)^\alpha K_\alpha(kx_0) \phi_0(\bvec{k}).
\end{equation}
For small $x_0$ this can be expanded as 
\begin{equation}
\label{phihat3} 
  \hat\phi(x) = \phi_0(\bvec{x}) +x_0^{2\alpha} c_\alpha \int d^dy\,
  \frac{\phi_0(\bvec{y})}{|\bvec{x-y}|^{d+2\alpha}} +
  \mathcal{O}\left(x_0^{2n},x_0^{2(\alpha+n)}\right). 
\end{equation}
Substituting eqn.\ \eqref{phihat3} into eqn.\ \eqref{onshell} one
obtains 
\begin{equation}
\label{Iwrong}
  I = -\frac12 \int d^dx\, \epsilon^{-2\alpha}
  \left(\frac{d}2-\alpha\right) \phi_0^2 - \frac{d}2 c_\alpha \int
  d^dx d^dy\,\frac{\phi_0(\bvec{x}) 
  \phi_0(\bvec{y})}{|\bvec{x-y}|^{d+2\alpha}}
  +\mathcal{O}\left(\epsilon^{2(n-\alpha)},\epsilon^{2n}\right).
\end{equation}
Obviously, the finite term in eqn.\ \eqref{Iwrong} does not agree
with eqn.\ \eqref{Ifin}, except for $d=2\alpha$, i.e.\ for $m=0$.  
The reason for the discrepancy is that the
first term on the right hand side of eqn.\ \eqref{onshell}, which is
purely divergent in the Dirichlet prescription, contributes to the
finite term, if eqn.\ \eqref{phihat2} is used. Ignoring this spurious
contribution (by including it into the counterterm), 
the finite terms coincide. Thus, one must accept
that counterterms are to be expressed in terms of $\phi_\epsilon$, not
$\phi_0$, which is the essence of the Dirichlet prescription.

\section{Divergent Terms for Gravity}
\label{anomaly}
\subsection{General Formalism}
The gravity action is given by \cite{Liu,Arutyunov} 
\begin{equation}
\label{gen:action}
  I = - \int_\epsilon d^{d+1}x \sqrt{\tilde g} \left[\tilde R
  +\frac{d(d-1)}{l^2} \right]
  + 2 \int d^dx \sqrt{g}\, \left[H+\frac{d-1}l\right],
\end{equation}
where the cosmological constant has been set equal to $2\Lambda
=-d(d-1)/l^2$. The last term in the boundary integral can be
considered as the first counterterm. 
As for our calculation of the finite part of the action
\cite{Mueck3} we use the time slicing formalism, which is summarized
in the appendix. Let us choose $\rho=X^0$ as time coordinate and use the
gauge 
\begin{equation}
\label{gen:gauge}
  n =\frac{l}{2\rho}, \qquad n^i=0.
\end{equation}
After isolating the leading behaviour of $g_{ij}$ for small $\rho$
(which can be found from the equation of motion) by defining 
\begin{equation}
\label{gen:ghatdef}
  g_{ij} = \frac1\rho \hat g_{ij}, 
\end{equation}
the equation of motion \eqref{sli:eqnmot} becomes
\begin{equation}
\label{gen:eqnmot1} 
  l^2 \hR_{ij} +(d-2) \hat g_{ij}' - 2\rho \hat g_{ij}'' + 2 \rho \hat
  g^{kl} \hat g_{ik}' \hat g_{lj}' - \hat g^{kl} \hat g_{kl}' \left(\rho
  \hat g_{ij}' -\hat g_{ij}\right) = 0.
\end{equation}
Here, $\hR_{ij}=R_{ij}$ is the Ricci tensor of the time slice
hypersurface. Raising an index with the metric $\hat g^{ij}$ we
realize that it is handy to define the quantity 
\begin{equation}
\label{gen:hdef}
  h^i_j = \hat g^{ik} \hat g_{kj}'.
\end{equation}
In fact, eqn.\ \eqref{gen:eqnmot1} becomes
\begin{equation}
\label{gen:eqnmot}
  l^2 \hR^i_j +(d-2) h^i_j +h \delta^i_j - \rho \left(2{h^i_j}' + h
  h^i_j \right) =0.
\end{equation}
Similarly, rewriting the constraints \eqref{sli:con1} and
\eqref{sli:con2} using eqns.\ \eqref{gen:gauge}, \eqref{gen:ghatdef}
and \eqref{gen:hdef} one obtains
\begin{equation}
\label{gen:con1}
  l^2 \hR + 2 (d-1) h +\rho \left( h^i_j h^j_i -h^2 \right) =0 
\end{equation}
and
\begin{equation}
\label{gen:con2}
  D_i h -D_j h^j_i = 0,
\end{equation}
respectively. 

In the AdS/CFT correspondence we have to calculate the on-shell value
of the action \eqref{gen:action} as a functional of prescribed
boundary values $\hat g_{ij}$, where the boundary is given by
$\rho=\epsilon$. First, the on-shell action is easily
found to be
\begin{equation}
\label{gen:onsh} 
  I(\epsilon) = \frac{d}l \int_\epsilon d\rho\, d^dx\, \sqrt{\hat g}\, 
  \rho^{-1-\frac{d}2} 
  +\frac2l \int d^dx \sqrt{\hat g}\, \epsilon^{-\frac{d}2} (\epsilon h -1).
\end{equation}
In order to find the singular terms in the limit $\epsilon\to0$, we
differentiate eqn.\ \eqref{gen:onsh} with respect to $\epsilon$,
leading to 
\begin{equation}
\label{gen:delI}
  \frac{\partial I}{\partial \epsilon} = \int d^d x \sqrt{\hat g}\,
  \epsilon^{-\frac{d}2} \left[ l \hR +\frac{d-1}l h \right].
\end{equation}
We have made use of the trace of the equation of motion
\eqref{gen:eqnmot} in order to simplify this expression. One can find
the singular terms by calculating $h$ from eqns.\ \eqref{gen:eqnmot},
\eqref{gen:con1} and \eqref{gen:con2} as a power series in $\epsilon$,
keeping only terms of order smaller than $\epsilon^\frac{d}2$. Thus, for
odd $d$, eqn.\ \eqref{gen:delI} contains only singular terms
proportional to powers $\epsilon^{-n+\frac12}$. On the other hand, for
even $d$, eqn.\ \eqref{gen:delI} contains a term proportional to
$1/\epsilon$, which yields a corresponding term proportional to $\ln
\epsilon$ in $I$. This logarithmic divergence is the source of the
Weyl anomaly in the regularized finite action. 
 
\subsection{$d=2$}
There is not really much to do for $d=2$. In fact, the divergent term
in eqn.\ \eqref{gen:delI} is obtained from the leading order solution
for $h$. Using the constraint \eqref{gen:con1} one finds
\begin{equation}
\label{d2:h}
  h = -\frac{l^2}2 \hR +\mathcal{O}(\epsilon).
\end{equation}
Hence, the divergent term in the action is
\begin{equation}
\label{d2:Idiv} 
  I_{div} = \ln\epsilon\; \frac{l}2 \int d^dx \sqrt{\hat g}\, \hR.
\end{equation}

\subsection{$d=4$}
Starting from the constraint \eqref{gen:con1} one finds
\begin{equation}
\label{d4:h1}
  h = -\frac16 \left[l^2 \hR+\epsilon \left(h^i_j h^j_i -h^2
  \right)\right].
\end{equation}
Here, the leading order behaviour of the term in parentheses is sufficcient.
The equation of motion \eqref{gen:eqnmot} gives
\[  h^i_j = -\frac{l^2}2 \left(\hR^i_j - \frac16 \delta^i_j \hR
  \right) +\mathcal{O}(\epsilon), \]
which in turn yields
\[ h^i_j h^j_i -h^2 = \frac{l^4}4 \left(\hR^i_j \hR^j_i - \frac13
  \hR^2 \right) +\mathcal{O}(\epsilon). \]
Hence, one finds 
\begin{equation}
\label{d4:Idiv} 
  I_{div} = - \int d^dx \sqrt{\hat g}\left[ \frac{l}{2\epsilon}\hR +
  \ln\epsilon\; \frac{l^3}8 \left(\hR^i_j \hR^j_i - \frac13
  \hR^2 \right)\right].
\end{equation}

\subsection{$d=6$}
The constraint \eqref{gen:con1} yields 
\begin{equation}
\label{d6:h1}
  h =-\frac1{10}\left[ l^2 \hR +\epsilon \left(h^i_j h^j_i-h^2
  \right)\right].
\end{equation}
We have to calculate the term in parentheses up to order
$\epsilon$. Starting from the equation of motion \eqref{gen:eqnmot} we
obtain
\[ h^i_j = -\frac{l^2}4 \left(\hR^i_j -\delta^i_j \frac{\hR}{10}
  \right) + \frac\epsilon4 \left[2{h^i_j}' +\frac{l^4\hR}{40}
  \left(\hR^i_j -\delta^i_j\frac{\hR}{10} \right) + \delta^i_j
  \frac1{10} \left( h^k_l h^l_k -h^2 \right) \right] +
  \mathcal{O}(\epsilon^2), \]
which in turn yields
\begin{equation}
\label{d6:h2}
  h^i_j h^j_i -h^2 = \frac{l^4}{16} \left( \hR^i_j \hR^j_i
  -\frac3{10} \hR^2 \right) - \frac\epsilon8 \left( 2l^2 \hR^j_i
  {h^i_j}' - \frac{l^2}5 \hR h' + \frac{15 l^6}{400} \hR \hR^i_j \hR^j_i 
  - \frac{29 l^6}{4000} \hR^3 \right) +\mathcal{O}(\epsilon^2).
\end{equation}
The quantities $h'$ and ${h^i_j}'$ can be found by differentiating the equation
of motion \eqref{gen:eqnmot} with respect to $\rho$, leading to
\begin{align}
\label{d6:hprime}
  h' &= -\frac18 \left(l^2 {\hR}' -h^2\right)+ \mathcal{O}(\epsilon),\\
\label{d6:hijprime}
  {h^i_j}' &= -\frac12 \left(l^2 {\hat {R^i_j}}' -\frac{l^2}8 {\hR}' 
  \delta^i_j
  + \frac18 \delta^i_j h^2 - h h^i_j \right) +\mathcal{O}(\epsilon).
\end{align}
The missing quantity ${\hat {R^i_j}}'$ is given by 
\begin{equation}
\label{d6:Rijprime}
  {\hat {R^i_j}}' = \frac12 \left( \hR^i_k h^k_j - \hR^k_j h^i_k
  \right) - \hR^{ik}_{\;\;jl} h^l_k +\frac12 D^i D_j h - \frac12
  D^k D_k h^i_j, 
\end{equation}
where we have used the constraint \eqref{gen:con2}. Taking the
trace of eqn.\ \eqref{d6:Rijprime} yields
\begin{equation}
\label{d6:Rprime}
  \hR' = - \hR^i_j h^j_i.
\end{equation}
Thus, substituting everything back into eqn.\ \eqref{d6:h2} we find
\begin{equation}
\label{d6:h3}
\begin{split}
  h^i_j h^j_i -h^2 &= \frac{l^4}{16} \left( \hR^i_j \hR^j_i
  -\frac3{10}\hR^2 \right) 
  -\frac{\epsilon l^6}{32} \left( \frac1{20} \hR D_k D^k \hR + 
  \frac15 \hR^i_j D^j D_i \hR -\frac12 \hR^i_j D_k D^k \hR^j_i\right.\\
  &\quad \left. - \hR^{ik}_{\;\;jl} \hR^j_i \hR^l_k
  +\frac12 \hR \hR^i_j \hR^j_i - \frac3{50} \hR^3 \right)
  +\mathcal{O}(\epsilon^2).
\end{split}
\end{equation}
Finally, substituting eqns.\ \eqref{d6:h1} and \eqref{d6:h3} into
eqn.\ \eqref{gen:delI} we obtain the result
\begin{equation}
\label{d6:Idiv}
\begin{split}
  I_{div} &= \int d^dx \sqrt{\hat g} \left[ -\frac{l}{4\epsilon^2} \hR
  + \frac{l^3}{32 \epsilon} \left( \hR^i_j \hR^j_i
  -\frac3{10}\hR^2 \right) 
  + \ln\epsilon\; \frac{l^5}{64} \left( \frac1{20} \hR D_k 
  D^k \hR \right. \right. \\
  &\quad\left.\left. + \frac15 \hR^i_j D^j D_i \hR -\frac12 \hR^i_j 
  D_k D^k \hR^j_i -\hR^{ik}_{\;\;jl} \hR^j_i \hR^l_k 
  + \frac12 \hR \hR^i_j \hR^j_i - \frac3{50} \hR^3
  \right) \right]. 
\end{split}
\end{equation}

\section{Conclusions}
In this paper, we first explained the regularization procedure for the
AdS/CFT correspondence. This was done using the example of a massive
scalar field. The regularization procedure involves considering a
family of surfaces as space time boundary, which tend towards the 
AdS horizon for
some limit $\epsilon\to0$. When using the cut-off, one must express
all counterterms in terms of the boundary values of the AdS fields at 
the cut-off boundary, not the asymptotic horizon value. Our example
demonstrates the importance of this step and thus shows that the
``Dirichlet prescription'' is the only entirely consistent one known
so far. 

Then, we calculated the divergent terms for AdS gravity for $d=2,4,6$
using the Dirichlet prescription.
We found agreement with earlier results, whose derivation did not
properly address the boundary value subtlety \cite{Henningson2}, or
used different techniques \cite{Hyun,Balasu,Emparan}. 
The fact that the subtlety does not affect the result should be
regarded as a coincidence, as in the other cases mentioned in the
introduction. In fact, we calculated some divergent terms for the
scalar field and found that they generically disagree for the correct 
and asymptotic boundary values -- even in the massless case, where the
finite terms coincide. 

As in our calculation of the finite term \cite{Mueck3}, which yields
the two-point function of CFT energy momentum tensors, the time
slicing formalism proves a valuable tool for the gravity part of the
AdS/CFT correspondence. Moreover, we found that the calculation was
greatly simplified by considering the
derivative of the action, eqn.\ \eqref{gen:delI}.

\section*{Acknowledgments}
This work was supported in part by a grant from NSERC. 
W.~M.\ is very grateful to Simon Fraser University for financial support.

\numberwithin{equation}{section}
\begin{appendix}
\section{Time Slicing Formalism}
\label{slicing}
Let us begin with a review of basic geometric relations for immersed
hypersurfaces \cite{Eisenhart}. Let a hypersurface be defined by the
functions $X^\mu(x^i)$, ($\mu=0,\ldots d$, $i=1,\ldots d$) and let
$\tilde g_{\mu\nu}$ and $g_{ij}$ be the metric tensors of the
imbedding manifold and the hypersurface, respectively. The tangents
$\partial_i X^\mu$ and the normal $N^\mu$ of the hypersurface satisfy
the following orthogonality relations: 
\begin{align}
\label{sli:induced}
  \tilde g_{\mu\nu} \partial_i X^\mu \partial_j X^\nu &= g_{ij},\\
\label{sli:orth}
  \partial_i X^\mu N_\mu &= 0,\\
\label{sli:norm}
  N_\mu N^\mu &=1.
\end{align}
We shall in the sequel use a tilde to label quantities relating to the
$d+1$ dimensional space time manifold and leave those relating to the
hypersurface unadorned. Moreover, we use the symbol $D$ to denote a
covariant derivative with respect to whatever indices may
follow. Then, there are the equations of Gauss and Weingarten, which
define the second fundamental form $H_{ij}$ of the hypersurface,
\begin{align}
\label{sli:gauss1}
  D_i\partial_j X^\mu &\equiv \partial_i\partial_j X^\mu + \tilde
  \Gamma^\mu_{\lambda\nu} \partial_i X^\lambda \partial_j X^\nu
  -\Gamma^k_{ij} \partial_k X^\mu = H_{ij} N^\mu,\\
\label{sli:weingarten}
  D_i N^\mu &\equiv \partial_i N^\mu +\tilde \Gamma^\mu_{\lambda\nu}
  \partial_i X^\lambda N^\nu = -H^j_i \partial_j X^\mu.
\end{align}
The second fundamental form describes the extrinsic curvature of the
hypersurface and is related to the intrinsic curvature by another
equation of Gauss,
\begin{align}
\label{sli:gauss2}
  \tilde R_{\mu\nu\lambda\rho} \partial_i X^\mu \partial_j X^\nu
  \partial_k X^\lambda \partial_l X^\rho &=R_{ijkl} + H_{il} H_{jk} -
  H_{ik} H_{jl}.\\
\intertext{Furthermore, it satisfies the equation of Codazzi,}
\label{sli:codazzi}
  \tilde R_{\mu\nu\lambda\rho} \partial_i X^\mu \partial_j X^\nu
  N^\lambda \partial_k X^\rho &= D_i H_{jk} -D_j H_{ik}.
\end{align}

In the time slicing formalism \cite{Wald,MTW} we consider the bundle
of immersed hypersurfaces defined by $X^0=const.$, whose tangent
vectors are given by $\partial_i X^0=0$ and $\partial_i X^\mu
=\delta_i^\mu$ ($\mu=1,\ldots d$). One conveniently splits up the
metric as (shown here for Euclidean signature)
\begin{align}
\label{sli:split}
  \tilde g_{\mu\nu} &= \begin{pmatrix} n_i n^i +n^2 & n_j \\
				n_i & g_{ij} \end{pmatrix},\\
\intertext{whose inverse is given by}
\label{sli:splitinv}
  \tilde g^{\mu\nu} &= \frac1{n^2} \begin{pmatrix} 1& -n^j \\
			-n^i & n^2 g^{ij} +n^i n^j \end{pmatrix}
\end{align}
and whose determinant is $\tilde g=n^2 g$. The quantities $n$ and
$n^i$ are called the lapse function and shift vector,
respectively. The normal vector $N^\mu$ satisfying eqns.\
\eqref{sli:orth} and \eqref{sli:norm} is given by 
\begin{equation}
\label{sli:normal}
  N_\mu = (-n,\bvec{0}), \qquad N^\mu= \frac1{n}(-1,n^i),
\end{equation}
where the sign has been chosen such that the normal points outwards on
the boundary ($n>0$ without loss of generality). Then, one obtains the
second fundamental form from the equation of Gauss \eqref{sli:gauss1}  
as 
\begin{equation}
\label{sli:Hij}
  H_{ij} = \frac1{2n} (g_{ij}' -D_i n_j -D_j n_i),
\end{equation}
where the prime denotes a derivative with respect to the time coordinate
($X^0$).  

The advantage of the time slicing formalism is that one removes the
diffeomorphism invariance in Einstein's equation by specifying the
lapse function $n$ and shift vector $n^i$ and thus obtains an equation of
motion as well as constraints for the physical degrees of freedom
$g_{ij}$. Consider Einstein's equation without matter fields,
\begin{equation}
\label{sli:einstein} 
  \tilde R_{\mu\nu} -\frac12 \tilde g_{\mu\nu} \tilde R = - \tilde
  g_{\mu\nu} \Lambda.
\end{equation} 
Multiplying it with $N^\mu N^\nu$ and using the equation of Gauss
\eqref{sli:gauss2} as well as the relation \eqref{sli:norm} one
obtains the first constraint,
\begin{equation} 
\label{sli:con1}
  R+H_{ij}H^{ij} -H^2 = 2\Lambda, 
\end{equation} 
where $H=H^i_i$. Similarly, multiplying with $N^\mu \partial_i X^\nu$,
using the equation of Codazzi \eqref{sli:codazzi} and the relation
\eqref{sli:orth} yields the second constraint,
\begin{equation}
\label{sli:con2}
  D_i H - D_j H^j_i =0.
\end{equation}
Finally, rewriting eqn.\ \eqref{sli:einstein} in the form
\begin{equation}
\label{sli:einstein2} 
  \tilde R_{\mu\nu} = \frac{2}{d-1}\tilde g_{\mu\nu} \Lambda
\end{equation}
and projecting out its tangential components we obtain the equation of
motion
\begin{equation} 
\label{sli:eqnmot}
  \tilde R_{ij} = \frac2{d-1} g_{ij} \Lambda.
\end{equation}
  
\end{appendix}


\begin{thebibliography}{99}
\bibitem{Gubser} S.~S.~Gubser, I.~R.~Klebanov and A.~M.~Polyakov,
  Phys.~Lett.~\textbf{B428}, 105 (1998), \texttt{hep-th/9802109}
\bibitem{Witten} E.~Witten, Adv.~Theor.~Math.~Phys.~\textbf{2}, 253
  (1998), \texttt{hep-th/9802150}
\bibitem{Mueck1} W.~M\"uck and K.~S.~Viswanathan, Phys.~Rev.~D
  \textbf{58}, 041901 (1998), \texttt{hep-th/9804035}
\bibitem{Freedman1} D.~Z.~Freedman, S.~D.~Mathur, A.~Matusis and
  L.~Rastelli, \emph{Correlation functions in the CFT$_d$/AdS$_{d+1}$
  correspondence}, \texttt{hep-th/9804058} 
\bibitem{Mueck2} W.~M\"uck and K.~S.~Viswanathan, Phys.\ Rev.\ D
  \textbf{58}, 106006 (1998), \texttt{hep-th/9805145}
\bibitem{Corely} S.~Corley, Phys.\ Rev.\ D \textbf{59}, 086003 (1998),
  \texttt{hep-th/9808184} 
\bibitem{Mueck3} W.~M\"uck and K.~S.~Viswanathan, \emph{The Graviton
  in the AdS-CFT correspondence: Solution via the Dirichlet boundary
  value problem}, \texttt{hep-th/9810151}
\bibitem{Henningson1} M.~Henningson and K.~Sfetsos, Phys.\ Lett.\
  \textbf{B431}, 63 (1998), \texttt{hep-th/9803251} 
\bibitem{Liu} H.~Liu and A.~A.~Tseytlin, Nucl.\ Phys.\ B \textbf{533},
  88 (1998), \texttt{hep-th/9804083}
\bibitem{AVolovich} A.~Volovich, J.\ High Energy Phys.\ \textbf{9},
  022 (1998), \texttt{hep-th/9809009}
\bibitem{Koshelev} A.~S.~Koshelev and O.~A.~Rytchkov, Phys.\ Lett.\
  \textbf{B450}, 368 (1999), \texttt{hep-th/9812238}   
\bibitem{Freedman2} E.~D'Hoker, D.~Z.~Freedman, S.~D.~Mathur,
  A.~Matusis and L.~Rastelli, \emph{Graviton exchange and complete
  4-point functions in the AdS/CFT correspondence}, \texttt{hep-th/9903196}
\bibitem{Henningson2} M.~Henningson and K.~Skenderis, J.\ High Energy
  Phys.\ \textbf{7}, 023 (1998)
\bibitem{Henningson3} M.~Henningson and K.~Skenderis, \emph{Holography
  and the Weyl Anomaly}, \texttt{hep-th/9812032}
\bibitem{Nojiri1} S.~Nojiri and S.~D.~Odintsov, Phys.\ Lett.\
 \textbf{B444}, 92 (1998), \texttt{hep-th/9810008}
\bibitem{Hyun} S.~Hyun, W.~T.~Kim and J.~Lee, Phys.\ Rev.\ D
  \textbf{59}, 084020 (1999), \texttt{hep-th/9811005}
\bibitem{Balasu} V.~Balasubramanian and P.~Kraus, \emph{A Stress
  Tensor for Anti-de Sitter Gravity}, \texttt{hep-th/9902121}
\bibitem{Nojiri2} S.~Nojiri and S.~D.~Odintsov, \texttt{hep-th/9903033}
\bibitem{Gonzalez} F.~Gonzalez-Rey, B.~Kulik and I.~Y.~Park,
  \emph{Non-renormalization of two and three Point Correlators of $N=4$
  SYM in $N=1$ Superspace}, \texttt{hep-th/9903094}
\bibitem{Emparan} R.~Emparan, C.~V.~Johnson and R.~C.~Myers,
  \emph{Surface Terms as Counterterms in the AdS/CFT Correspondence}, 
  \texttt{hep-th/9903238}
\bibitem{Nishi} M.~Nishimura and Y.~Tanii, \emph{Super Weyl Anomalies
  in the AdS/CFT Correspondence}, \texttt{hep-th/9904010}
\bibitem{Arutyunov} G.~E.~Arutyunov and S.~A.~Frolov, Nucl.\ Phys.\ B
  \textbf{544}, 576 (1999), \texttt{hep-th/9806216}
\bibitem{Eisenhart} L.~P.~Eisenhart, \emph{Riemannian Geometry},
  Princeton University Press (1964)
\bibitem{Wald} R.~M.~Wald, \emph{General Relativity}, University of
  Chicaco Press (1984) 
\bibitem{MTW} C.~W.~Misner, K.~S.~Thorne and J.~A.~Wheeler,
  \emph{Gravitation}, Freeman, San Francisco (1973)
\end{thebibliography}
\end{document}